\documentclass[doublecol]{epl2} 

\usepackage{epsfig,amsmath,amssymb} 

\title{The quantum $J_{1}$--$J_{1}'$--$J_{2}$ spin-$1$ Heisenberg model: 
Influence of the interchain coupling on the ground-state magnetic ordering in 2D}
\shorttitle{The quantum $J_{1}$--$J_{1}'$--$J_{2}$ spin-$1$ isotropic Heisenberg model} 

\author{R. F. Bishop\inst{1,2} \and P. H. Y. Li\inst{1,2} \and R. Darradi\inst{3} \and J. Richter\inst{3}}
\shortauthor{R. F. Bishop \etal}

\institute{                    
  \inst{1} School of Physics $\&$ Astronomy, Schuster Building, The University of Manchester, Manchester, M13 9PL, UK\\
  \inst{2} School of Physics $\&$ Astronomy, University of Minnesota, 116 Church St. SE, Minneapolis, Minnesota 55455, USA\\
  \inst{3} Institut f\"ur Theoretische Physik, Universit\"at Magdeburg, 39016 Magdeburg, Germany
}


\abstract{We study the phase diagram of the isotropic $J_{1}$--$J_{1}'$--$J_{2}$ Heisenberg
model for spin-$1$ particles on an anisotropic square lattice, 
using the coupled cluster method.  We find no evidence for an intermediate phase between the N\'{e}el and
stripe states, as compared with all previous results 
for the corresponding spin-$1/2$ case.  However, we find a quantum tricritical point at 
$J_{1}'/J_{1} \approx0.66 \pm 0.03$, $J_{2}/J_{1} \approx0.35\pm0.02$, where a line 
of second-order phase transitions between the quasi-classical N\'{e}el and stripe-ordered phases 
(for $J_{1}'/J_{1} \lesssim 0.66$) meets a line of first-order phase transitions between the same two 
states (for $J_{1}'/J_{1} \gtrsim 0.66$).}  
                      
\begin{document}

\maketitle

Over the last 20 years or so much theoretical 
effort~\cite{Ch:1988,Dag:1989,Ri:1993,Schulz:1996,Bi:1998_PRB58,Ca:2001,Siu:2001,Sin:2003} 
has been expended on the $J_{1}$--$J_{2}$ model in which the spins situated on the sites of a 
two-dimensional (2D) square lattice interact via competing isotropic Heisenberg interactions 
between the nearest-neighbour ($J_{1}$) and next-nearest-neighbour ($J_{2}$) pairs.  The exchange 
bonds $J_{1} > 0$ promote antiferromagnetic order, while the $J_{2} > 0$ bonds act to frustrate 
or compete with this order.  Such frustrated quantum magnets continue to be of interest because 
of the possible spin-liquid and other such novel phases that they can exhibit (see, e.g., 
Ref.~\cite{Le:2007}).  

The syntheses of compounds that can be closely described by the spin-1/2 version of the model, 
such as Li$_{2}$VO(Si,Ge)O$_{4}$~\cite{Mel:2000} and VOMoO$_{4}$~\cite{Car:2002} have further 
fuelled theoretical interest.  It is now widely accepted that the 
spin-1/2 $J_{1}$--$J_{2}$ model on the 2D square lattice has a ground-state phase diagram 
showing two phases with quasi-classical long-range order (LRO) (viz., a N\'{e}el-ordered phase at small values 
of $J_{2}/J_{1}$ and a collinear stripe-ordered phase at large values of $J_{2}/J_{1}$), 
separated by an intermediate quantum paramagnetic phase without magnetic LRO in the parameter 
regime $\alpha^{1}_{c} < \alpha < \alpha^{2}_{c}$, where $\alpha \equiv J_{2}/J_{1}$ and 
$\alpha^{1}_{c} \approx 0.4, \alpha^{2}_{c} \approx 0.6$.  Furthermore, it has 
been argued recently that the quantum phase transition
between the quasi-classical N\'eel phase and the quantum paramagmetic phase  
present in the 2D $J_1$--$J_2$ model is not described by a Ginzburg-Landau
type critical theory, but rather may exhibit 
a deconfined quantum critical point~\cite{Sen:2004}.  Other authors~\cite{Sir:2006} have 
argued that the transition is not of this second-order type due to the deconfinement of 
the fractionally-charged spinons, but is rather a (weakly) first-order transition between 
the N\'{e}el phase and a valence-bond solid phase with columnar dimerisation.

Such frustrated quantum magnets often have ground states that are macroscopically degenerate.  
This feature leads naturally to an increased sensitivity of the underlying Hamiltonian to 
the presence of small perturbations.  In particular, the presence of anisotropies in real 
systems that are well characterised by the $J_{1}$--$J_{2}$ model, either in spin space or 
in real space, naturally raises the issue of how robust are the properties of the $J_{1}$--$J_{2}$ 
model against any such perturbations.  There have been several recent studies addressing 
this question.  For example, in the case of spin anisotropies, generalizations of the 
$J_{1}$--$J_{2}$ model have been studied for the spin-1/2 case, in which either the 
frustrating next-nearest-neighbour interaction or the nearest-neighbour interaction is 
anisotropic~\cite{Ro:2004,Via:2007}.  

In the alternative case of real-space anisotropies, for example, a recent 
study~\cite{Schm:2006} investigated the effects of including an interlayer 
coupling ($J_{\perp}$) for the spin-half $J_{1}$--$J_{2}$ model on a stacked square 
lattice.  In a previous paper of our own~\cite{Bi:2007_j1j2j3_spinHalf} we moved 
instead in the direction of one-dimensionality by investigating a spin-half spatially 
inhomogeneous $J_{1}$--$J_{1}'$--$J_{2}$ model in which the nearest-neighbour bonds on 
the square lattice differ for the intrachain ($J_{1}$) and interchain ($J_{1}'$) cases.  
The model can thus be viewed as parallel ($J_{1}$) chains, coupled by nearest-neighbour 
($J_{1}'$) and next-nearest-neighbour ($J_{2}$) interactions that frustrate each other.  
We found the surprising and novel result that for the spin-1/2 case there exists a 
quantum triple point below which there is a second-order phase transition between the 
quasi-classical N\'{e}el and stripe-ordered phases with magnetic LRO, whereas 
only above this point are these two phases separated by the intermediate magnetically 
disordered liquid-like phase seen in the pure spin-1/2 $J_{1}$--$J_{2}$ model (i.e., 
at $J_{1}' = J_{1}$).  The quantum triple point was found to occur at 
$J_{1}'/J_{1} \approx 0.60 \pm 0.03, J_{2}/J_{1} \approx 0.33 \pm 0.02$. 

In the present work we extend the study of the $J_{1}$--$J_{1}'$--$J_{2}$ model to 
consider the spin-1 case, which is computationally more challenging than the previous 
spin-1/2 case.  As in the previous case we again use the much-studied 
coupled cluster method (CCM).  Our main rationale for the present study is 
that one knows in general that the spin quantum number can play an important and 
highly non-trivial role in these strongly correlated magnetic-lattice systems, which 
often exhibit rich and interesting phase scenarios due to the interplay between 
the quantum fluctuations and the competing interactions present.  The strength 
of the quantum fluctuations can be tuned either by introducing spin-anisotropy 
terms in the Hamiltonian~\cite{Da:2004} or by varying the spin quantum 
number $s$~\cite{Da:2005_JPhy_17}.  

While the general trend is that as $s$ is increased the effects of 
quantum fluctuations reduce, one also knows that there can be significant 
deviations from it.  A particularly well-known example is the since-confirmed 
prediction of Haldane that integer-spin systems on the linear chain would have a 
nonzero excitation energy gap, whereas half-odd-integer spin systems would be 
gapless~\cite{Ha:1983}.  Indeed, such deviations from general trends provide one 
of the main reasons why quantum spin-lattice problems still maintain such an 
important role in the general investigation of quantum phase transitions.

In this context we note that the recent discovery of superconductivity with a 
transition temperature at $T_c \approx 26\,$K in the layered iron-based compound LaOFeAs, 
when doped by partial substitution of the oxygen atoms by fluorine atoms~\cite{KWHH:2008}, 
has been followed by the rapid discovery of superconductivity at even higher values of $T_c$ 
($\gtrsim 50\,$K) in a broad class of similar quaternary compounds.  Enormous interest has 
thereby been engendered in this class of materials.  Of particular relevance to the present 
work are the very recent first-principles calculations~\cite{MLX:2008} showing that the undoped 
parent precursor material LaOFeAs is well described by the spin-1 $J_{1}$--$J_{2}$ model on the 
square lattice with $J_1 > 0$, $J_2 > 0$, and $J_{2}/J_{1} \approx 2$.

Returning to our present system, we note that while the $s=1/2$ version of the $J_{1}$--$J_{1}'$--$J_{2}$ 
model under discussion has been studied by various
groups~\cite{Bi:2007_j1j2j3_spinHalf,Ne:2003,Si:2004,Star:2004,Mo:2006}, very few calculations 
have been performed on the $s=1$ case up till now.  An exception is the two-step density-matrix 
renormalisation group study of Moukouri~\cite{Mo:2006,Mo:2006_b} that we discuss later in 
our concluding remarks.  It has also been observed that quantum fluctuations
can destabilize the ordered classical ground state (GS), even for values $s>1/2$, for
large enough values of the frustration~\cite{Ch:1988,Kr:2006}.  

The model itself comprises a set of $N \rightarrow \infty$ spin-1 particles on a spatially 
anisotropic square lattice interacting  via isotropic Heisenberg couplings, but with 
three kinds of exchange bonds.  Its Hamiltonian is given by 
\begin{eqnarray}
H &=& J_{1}\sum_{i,l}\vect{s}_{i,l}\cdot \vect{s}_{i+1,l} + J_{1}'\sum_{i,l}\vect{s}_{i,l}\cdot \vect{s}_{i,l+1}\nonumber\\
 &+& J_{2}\sum_{i,l}(\vect{s}_{i,l}\cdot \vect{s}_{i+1,l+1} + \vect{s}_{i+1,l} \cdot \vect{s}_{i,l+1}),  \label{H} 
\end{eqnarray}
where the index ($i,l$) labels the $x$ (row) and $y$ (column) components of the lattice sites.
The exchange constant $J_{1}$ (which we henceforth set to 1) measures the intrachain bond 
strength along the row direction, while $J_{1}'$ and $J_{2}$ are the transverse (column) 
and diagonal interchain couplings respectively.  
The model retains the basic physics of the $J_1$--$J_2$ model (that is recovered when $J_{1}'=J_{1}$), 
and has two types of classical GS, namely, the
N\'{e}el ($\pi,\pi$) state and stripe states (columnar stripe
($\pi,0$) and row stripe ($0,\pi$)).  There is clearly a symmetry
under the interchange of rows and columns, $J_{1} \rightleftharpoons
J_{1}'$, which implies that we need only consider the range of
parameters with $J_{1}'<J_{1}$.  The (first-order) classical phase transition 
occurs at the point of maximal frustration, \(J^{c}_{2}=J_{1}'/2,\; \forall J_{1}>J_{1}'\).

The CCM (see, e.g., Refs.~\cite{Bi:1991,Bi:1998,Fa:2004} and references 
cited therein) employed here is one of the most powerful and 
most versatile modern techniques in quantum many-body theory.  It has been successfully applied  
to various quantum magnets (see 
Refs.~\cite{Ze:1998,Kr:2000,Fa:2002,Fa:2004,Schm:2006,Zi:2008} and references cited therein).
The CCM is particularly appropriate for studying frustrated systems, for which the main alternative
methods are often only of limited usefulness.  For example, quantum Monte Carlo techniques 
are particularly plagued by the sign problem for such systems, and the exact diagonalisation method 
is restricted in practice, particularly for $s>1/2$, to such small lattices that 
it is often insensitive to the details of any subtle phase order present.

We now briefly describe the CCM means to solve the ground-state (gs) Schr\"{o}dinger 
ket and bra equations, $H|\Psi\rangle = E|\Psi\rangle$ 
and $\langle\tilde{\Psi}|H=E\langle\tilde{\Psi}|$ respectively (and see 
Refs.~\cite{Bi:1991,Bi:1998,Kr:2000,Ze:1998,Fa:2002,Fa:2004} for further details).  
The first step in implementing the CCM is always to choose a model state $|\Phi\rangle$ 
on top of which to incorporate 
later in a systematic fashion the multispin correlations contained in the exact ground 
states $|\Psi\rangle$ and $\langle\tilde{\Psi}|$.  More specifically, the CCM employs the 
exponential ansatz, $|\Psi\rangle=$e$^{S}|\Phi\rangle$ 
and $\langle\tilde{\Psi}|=\langle\Phi|\tilde{\cal S}$e$^{-S}$.  The correlation operator 
$S$ is expressed as $S = \sum_{I\neq0}{\cal S}_{I}C^{+}_{I}$ and its counterpart is 
$\tilde{S} = 1 + \sum_{I\neq0}\tilde{\cal S}_{I}C^{-}_{I}$.  The operators 
$C^{+}_{I} \equiv (C^{-}_{I})^{\dagger}$, with $C^{+}_{0} \equiv 1$, have the property 
that $\langle\Phi|C^{+}_{I} = 0\,;\, \forall I \neq 0$.  They form a complete 
set of multispin creation operators
with respect to the model state $|\Phi\rangle$.  The ket- and bra-state correlation 
coefficients $({\cal S}_{I}, \tilde{{\cal S}_{I}})$
are calculated by requiring the gs energy expectation value
$\bar{H} \equiv \langle\tilde{\Psi}|H|\Psi\rangle$
to be a minimum with respect to each of them.
This immediately yields the coupled set of equations
$\langle \Phi|C^{-}_{I}\mbox{e}^{-S}H\mbox{e}^{S}|\Phi\rangle = 0$ and
$\langle\Phi|\tilde{S}(\mbox{e}^{-S}H\mbox{e}^{S} - E)C^{+}_{I}|\Phi\rangle =
0\,;\, \forall I \neq 0$, which we solve in practice for the correlation coefficients
$({\cal S}_{I}, \tilde{{\cal S}_{I}})$ within specific truncation schemes described below, 
by making use of parallel computing routines~\cite{ccm}. 

In order to treat each lattice site on an equal footing we  
perform a mathematical rotation of the local spin axes on each lattice site such
that every spin of the model state aligns along its negative $z$-axis.  Henceforth 
our description of the spins is given wholly in terms of these locally defined spin
coordinate frames.  In particular, the multispin creation operators may be written as 
\(C^{+}_{I}\equiv s^{+}_{i_{1}} s^{+}_{i_{2}} \cdots s^{+}_{i_{n}}\),
in terms of the locally defined spin-raising operators $s^{+}_{i} \equiv s^{x}_{i} + s^{y}_{i}$
on lattice sites $i$.  Having solved for the multispin cluster correlation coefficients 
$({\cal S}_{I}, \tilde{{\cal S}_{I}})$ as described above, we may then calculate the
gs energy $E$ from the relation 
$E=\langle\Phi|\mbox{e}^{-S}H\mbox{e}^{S}|\Phi\rangle$, and
the gs staggered magnetisation $M$ from the relation
$M \equiv -\frac{1}{N} \langle\tilde{\Psi}|\sum_{i=1}^{N}s^{z}_{i}|\Psi\rangle$
which holds in the rotated spin coordinates.  

Although the CCM formalism is clearly exact if a complete set of multispin
configurations $\{I\}$ with respect to the model state $|\Phi\rangle$ 
is included in the calculation of the correlation operators
$S$ and $\tilde{S}$, in practice it is necessary to use systematic approximation 
schemes to truncate them to some finite subset.  In our earlier paper on the $s=1/2$ 
version of the present model~\cite{Bi:2007_j1j2j3_spinHalf}, 
we employed, as in our previous
work~\cite{Fa:2004,Ze:1998,Kr:2000,Fa:2002,Schm:2006},
the localised LSUB$n$ scheme 
in which all possible multi-spin-flip correlations over different locales
on the lattice defined by $n$ or fewer contiguous lattice sites are retained.

However, we note that the number of fundamental LSUB$n$ configurations for $s=1$
becomes appreciably higher than for $s=1/2$, since each spin on each site $i$
can now be flipped twice by the spin-raising operator $s^{+}_{i}$.
Thus, for the $s=1$ model it is more practical, but equally systematic, to use
the alternative SUB$n$--$m$ scheme, in which all correlations involving up
to $n$ spin flips spanning a range of no more than $m$ adjacent lattice sites 
are retained~\cite{Fa:2004,Fa:2001}.  We then set $m=n$, and hence employ the 
so-called SUB$n$--$n$ scheme.  More generally, the LSUB$m$ scheme is 
thus equivalent to the SUB$n$--$m$ scheme for $n=2sm$ for particles of spin $s$.  
For $s=1/2$, LSUB$n\equiv$ SUB$n$--$n$;
whereas for $s=1$, LSUB$n\equiv$ SUB2$n$--$n$.  The numbers of such 
fundamental configurations (viz., those that are distinct
under the symmetries of the Hamiltonian and of the model state $|\Phi\rangle$)
that are retained for the N\'{e}el and stripe states of the current $s=1$ model at
various SUB$n$--$n$ levels are shown in Table~\ref{FundConf_spin1_SUBnn}. 

\begin{table}[!tbp]
\begin{center}
\caption[Number of fundamental SUB$n$--$n$ configurations, $J_{1}$--$J'_{1}$--$J_{2}$ model, 
s=1]{Number of fundamental configurations ({$\sharp$ f\@.c\@}) for the SUB$n$--$n$ ($n=\{2,4,6,8\}$) 
scheme for the N\'{e}el and stripe columnar states, for the spin-1 $J_{1}$--$J'_{1}$--$J_{2}$ model.}
\label{FundConf_spin1_SUBnn}
\vskip0.5cm
\begin{tabular}{|c|c|c|} \hline\hline
{Method} & \multicolumn{2}{c|}{$\sharp$ f\@.c\@} \\ \hline
SUB$n$--$n$ & N\'{e}el & stripe \\ \hline
SUB$2$--$2$ & 2 & 1 \\ \hline
SUB$4$--$4$ & 28 & 21 \\ \hline
SUB$6$--$6$ & 744 & 585 \\ \hline
SUB$8$--$8$ & 35629 & 29411 \\ \hline\hline
\end{tabular} 
\end{center}
\end{table}

Although we never need to perform any finite-size scaling, since all CCM approximations are 
automatically performed from the outset in the $N \rightarrow \infty$ limit, we do need as a 
last step to extrapolate to the $n \rightarrow \infty$ limit in the truncation index $n$.
We use the same well-tested scaling laws as for the $s=1/2$ model for the gs energy per spin $E/N$ and 
the gs staggered magnetisation $M$,
\begin{equation}
E/N=a_{0}+a_{1}n^{-2}+a_{2}n^{-4}\;,  \label{Extrapo_E}
\end{equation} 
\begin{equation}
M=b_{0}+n^{-0.5}\left(b_{1}+b_{2}n^{-1}\right)\;.  \label{Extrapo_M}
\end{equation} 
We report below on two separate sets of CCM calculations for this model, 
for given parameters $(J_{1} \equiv 1, J_{1}', J_{2})$, based respectively
on the N\'{e}el state and the stripe state as the model state $|\Phi\rangle$.
In each case we have 4 calculated data points to fit the 3 unknown parameters in 
eqs.\ (\ref{Extrapo_E}) and (\ref{Extrapo_M}).  Thus, we present below our final 
results $a_0$ for $E/N$ and $b_0$ for $M$, from fitting to the above 
schemes with the SUB$n$--$n$ solutions for $n=\{2,4,6,8\}$ as input.
We note that for the corresponding $s=1/2$ model we could perform 
LSUB$n\equiv$ SUB$n$--$n$ approximation calculations for $n=\{2,4,6,8,10\}$.  
This enabled us to perform extrapolations using the sets 
$n=\{2,4,6,8\}$ and $n=\{2,4,6,8,10\}$ as well as the preferred set 
$n=\{4,6,8,10\}$.  Gratifyingly, all sets yielded very similar extrapolated 
results, even near phase boundaries and the quantum triple point, which gave us 
great confidence in the accuracy and robustness of our extrapolation scheme.

\begin{figure}[!tbp]
\begin{center}
\epsfig{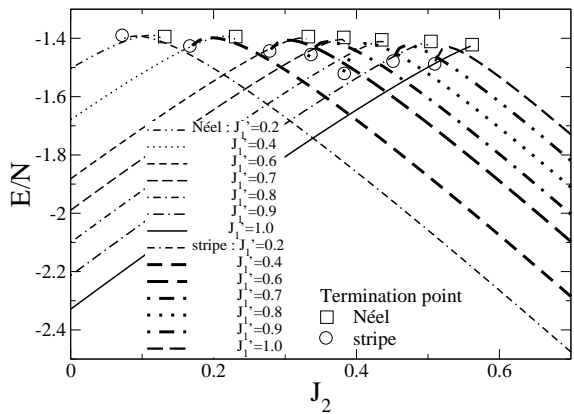}
\caption{Extrapolated CCM SUB$n$--$n$ results for the gs energy per spin, $E/N$, of the 
spin-1 $J_{1}$--$J_{1}'$--$J_{2}$ model (with $J_1=1$), for 
$J_{1}'=0.2,0.4,0.6,0,7,0.8,0.9,1.0$.  The SUB$n$--$n$ results are extrapolated 
in the limit $n \rightarrow \infty$ using the set $n=\{2,4,6,8\}$.}
\label{E_spin1}
\end{center}
\end{figure}

Figure~\ref{E_spin1} shows the gs energy per spin as a
function of $J_{2}$ for various values of $J_{1}'$, 
extrapolated from the raw CCM data as discussed above.
The raw SUB$n$--$n$ data terminate at some particular 
values.  This occurs for the CCM curves
based on both the N\'{e}el state and the stripe state as the model
state $|\Phi\rangle$.  Such a termination point arises due to the solutions of the
CCM equations becoming complex at this point, beyond which there
exist two branches of complex-conjugate solutions~\cite{Fa:2004}.  In the region
where the solution reflecting the true physical situation is real,
there actually also exists another real solution.  However, only
the (shown) upper branch of these two solutions reflects the true physical
situation, whereas the lower branch does not.  The physical branch is 
easily identified in practice as the one which becomes
exact in some known (e.g., perturbative) limit.
This physical branch then meets the corresponding unphysical branch
at some termination point beyond which no real solutions exist.
The termination points shown in fig.~\ref{E_spin1} are the extrapolated 
$n \rightarrow \infty$ termination
points and are evaluated using data only up to the highest level of the
CCM approximation schemes used here, namely SUB8--8 for the $s=1$ model.  The SUB$n$--$n$ 
termination points are also reflections of
phase transitions in the real system, as we discuss more fully below.

The maxima of the extrapolated gs energy curves are close to the 
corresponding classical transition points at $J^{c}_{2}=0.5J_{1}'$.  The extrapolated gs 
energy curves of the N\'{e}el and stripe states meet smoothly with the same slope 
for $J_{1}'\lesssim 0.66 \pm 0.03$.  This behaviour is indicative of a second-order 
phase transition.  By contrast, for $J_{1}' \gtrsim 0.66 \pm 0.03$ the behaviour is 
typical of a first-order phase transition where the curves cross with a discontinuity 
in the slope.  A comparison of fig.~\ref{E_spin1} for the present $s=1$ model with 
fig.\ 2 of Ref.~\cite{Bi:2007_j1j2j3_spinHalf} for the $s=1/2$ model clearly shows 
the distinct differences between the two cases.  Thus, for the $s=1/2$ case each gs 
energy curve for the N\'{e}el state for values
$J_{1}' \gtrsim 0.60$ terminates before it can reach 
the corresponding gs energy curve for the stripe state within the region that reflects the true 
physical situation (viz., where the calculated staggered magnetisation is positive), 
indicating the opening up of an
intermediate quantum phase between the N\'{e}el and stripe phases.  
By contrast, for the $s=1$ case, the gs energy curves of the N\'{e}el state for all values of 
$J_1'$ cross or meet the gs energy curves of the stripe state within 
the same physical region described above. 
  
\begin{figure}[!tbp]
\begin{center}
\epsfig{file=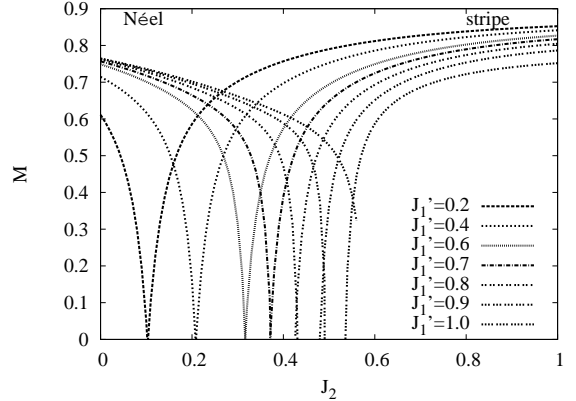,width=5.5cm,angle=270}
\caption{Extrapolated CCM SUB$n$--$n$ results for the gs staggered magnetisation, $M$, of the 
spin-1 $J_{1}$--$J_{1}'$--$J_{2}$ model (with $J_1=1$), for 
$J_{1}'=0.2,0.4,0.6,0,7,0.8,0.9,1.0$.  The SUB$n$--$n$ results are extrapolated 
in the limit $n \rightarrow \infty$ using the set $n=\{2,4,6,8\}$.}
\label{M_spin1}
\end{center}
\end{figure}

Figure~\ref{M_spin1} 
shows our corresponding extrapolated results for the gs staggered
magnetisation $M$.  The quantum phase
transition or critical point ($M_{\triangle_{c}}$) marking the end of either the quantum N\'{e}el state 
or the quantum stripe state for a given value of $J_{1}'$ is first determined by 
calculating the order parameter $M$ to obtain the value of
$J_{2}$ where $M$ vanishes.  
However, as seen in fig.~\ref{M_spin1}, there also occur
cases where the order parameters curves for the two states cross before their 
respective vanishing points.  In such cases we take the crossing point  
to indicate the phase boundary between the 
quantum N\'{e}el and quantum stripe states.  
Thus, our definition of the quantum critical point is the point
where there is an apparent first-order phase transition between 
the two states or where the order parameter
vanishes, whichever occurs first.  A fuller discussion of this 
former criterion and its relation to the stricter energy 
crossing criterion is given elsewhere~\cite{Schm:2006}.  

We note particularly the result for this $s=1$ model that
the order parameter $M$ curves for both the quantum N\'{e}el and stripe phases 
with the same value of $J_{1}'$ go to zero smoothly at the same point, 
for all values of $J_{1}'\lesssim 0.66 \pm 0.03$. 
We emphasize that this cannot be accidental since it occurs for a large 
number of essentially independent calculations over a wide parameter range.  
We also take this as further strong evidence for the accuracy and robustness 
of our extrapolation scheme.
Thus, in this regime we have 
behaviour typical of a second-order phase transition between the quantum 
N\'{e}el and stripe phases.  Furthermore, the transition occurs at a 
value of $J_2$ very close to the classical transition point 
at $J^{c}_{2}=0.5J_{1}'$.  Conversely, for values of $J_{1}' \gtrsim 0.66 \pm 0.03$, 
the order parameters $M$ of the two states meet at a
finite value, as is typical of a first-order transition.  

\begin{figure}[!tbp]
\begin{center}
\epsfig{file=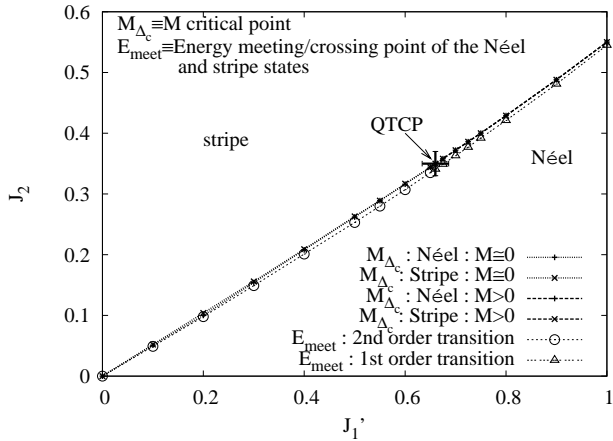,width=6cm,angle=270}
\caption{Extrapolated CCM SUB$n$--$n$ results for the gs phase diagram of the spin-1 $J_{1}$--$J_{1}'$--$J_{2}$ 
model (with $J_1=1$), showing a quantum tricritical point (QTCP).  The 
SUB$n$--$n$ results are extrapolated in the limit $n \rightarrow \infty$ using the set $n=\{2,4,6,8\}$.} 
\label{spin1_phase}
\end{center}
\end{figure}

Figure~\ref{spin1_phase} 
shows the zero-temperature phase diagram of the spin-$1$
$J_{1}$--$J_{1}'$--$J_{2}$ model on the square lattice, 
as obtained from our extrapolated 
results for both the gs energy and the gs order parameter $M$.  
Unlike the spin-1/2 case there is no sign at all of any intermediate 
disordered phase for any values of the parameters $J_{1}'$ or $J_{2}$ (for $J_{1}=1$).  
Hence, in this respect it is just like the classical case.  However, unlike the classical case, 
there now appears to be a quantum tricritical point (QTCP) at  
$J_{1}'\approx0.66 \pm 0.03$, $J_{2}\approx0.35\pm0.02$, where a line of second-order phase 
transitions between the quasi-classical N\'{e}el and stripe-ordered phases 
(for $J_{1}' \lesssim 0.66$) meets a line of first-order phase 
transitions between the same two states (for $J_{1}' \gtrsim 0.66$).  We note that the 
behaviour of both the order parameter (which goes to zero smoothly at the same 
point for both N\'{e}el and stripe phases below the QTCP, but which goes to a 
nonzero value above it) and of the gs energy curves for the two phases 
(which meet smoothly with the same slope below the QTCP, but which cross with a 
discontinuity in slope above it) tell exactly the same story, as observed in other
similar cases~\cite{Schm:2006}.  

Thus, there is no evidence from our work for an intermediate phase (for larger values 
of $J_{1}'/J_{1}$) for the $s=1$ case, by contrast with the $s=1/2$ case from our own 
previous results~\cite{Bi:2007_j1j2j3_spinHalf} (from which we found an intermediate 
phase without magnetic LRO for $J_{1}'/J_{1} \gtrsim 0.60 \pm 0.03$) and those of other 
groups~\cite{Si:2004,Star:2004,Mo:2006}.  For the spin-1/2 case this intermediate 
magnetically disordered phase was shown by us~\cite{Bi:2007_j1j2j3_spinHalf} to exist 
for the pure $J_{1}$--$J_{2}$ model (i.e., with $J_{1}'=J_{1}$) in the parameter 
range $\alpha^{1}_{c} < \alpha < \alpha^{2}_{c}$ for $\alpha \equiv J_{2}/J_{1}$, 
where $\alpha^{1}_{c} \approx 0.44 \pm 0.01$, $\alpha^{2}_{c} \approx 0.59 \pm 0.01$, 
in full agreement with the accepted values.  By contrast, for the $s=1$ model presented 
here we find instead a QTCP at $J_{1}'/J_{1} \approx 0.66 \pm 0.03$, $J_{2}/J_{1} \approx 0.35 \pm 0.02$.

For the case of the isotropic $J_1$--$J_2$ model lowest-order (or linear) spin-wave 
theory (LSWT)~\cite{Ch:1988} predicts that quantum fluctuations can destabilize the 
classical GS with LRO even at large values of the spin quantum number $s$, for values of 
$\alpha \equiv J_{2}/J_{1}$ around 0.5.  For the spin-1/2 case the range of values 
$\alpha^{1}_{c} < \alpha < \alpha^{2}_{c}$, for which a magnetically-disordered phase 
occurs is predicted by LSWT to be given by $\alpha^{1}_{c} \approx 0.38$, 
$\alpha^{2}_{c} \approx 0.52$.  For the corresponding spin-1 case LSWT predicts a 
much narrower, but still non-vanishing, strip of disordered intermediate phase with 
$\alpha^{1}_{c} \approx 0.47$, $\alpha^{2}_{c} \approx 0.501$.  However, in an 
important paper, Igarashi~\cite{Ig:1993} has shown explicitly for the spin-1 case, 
by going to higher-order terms in the $1/s$ power expansion of spin-wave theory 
(SWT), that no predictions based on LSWT (or SWT more generally) can be relied upon 
for values $J_{2}/J_{1} \gtrsim 0.4$ since the series seems to diverge in this region, 
with second-order terms becoming exceptionally large.  Igarashi also showed that the 
higher-order correction terms to LSWT act to make the N\'{e}el-ordered phase more 
stable than LSWT would predict.  We note too that Read and Sachdev~\cite{RS:1991}, using 
a large-$N$ expansion technique based on symplectic Sp($N$) symmetry, which can 
itself be regarded as akin to a $1/s$ expansion, have also found for the isotropic 
$J_1$--$J_2$ model an intermediate phase (with valence-bond solid order) for 
smaller values of $s$, which disappears for larger values of $s$ where they 
find instead a first-order transition between the N\'{e}el and stripe phases.  All 
of these results for the pure $J_1$--$J_2$ model are in accord with ours.

Naturally, one can also validly argue that what we have observed as a 
continuous (second-order) transition below the QTCP (i.e., 
for $J_{1}'/J_{1} \lesssim 0.66$) might actually be a very weak first-order 
transition, which would thereby still comply with the Landau symmetry criterion 
of the standard Ginzburg-Landau theory of phase transitions and critical phenomena.  
Our completely independent sets of CCM calculations based on the two quasi-classical  
phases can never entirely exclude this possibility.  However, our results from 
figs.~\ref{E_spin1} and~\ref{M_spin1} show clearly that the data below the QTCP 
are really only consistent with a transition which, if it is not second-order is at 
best very weakly first-order for all values of $J_{1}'/J_{1}$ below the QTCP.  In 
this context it is relevant to mention again that it has also been argued by 
others~\cite{Sen:2004} that for the equivalent spin-1/2 model the phase transition 
between the N\'{e}el state and the intermediate paramagnetic state (which has been 
argued by those authors to be a valence-bond solid state) is also second-order and 
hence not described by standard Ginzburg-Landau critical theory.  Again, for the 
spin-1/2 pure $J_{1}$--$J_{2}$ model the standard view is that the quantum phase 
transition between the striped and magnetically disordered intermediate phases 
is first-order, and there is no discussion in the literature of deconfined quantum 
criticality for this transition.  One might argue, on similar grounds, that for 
our $s=1$ model a first-order transition for the stripe phase might be more likely 
than for the N\'{e}el phase.  We stress again, however, that our own results do 
indicate a direct second-order transition between these two phases below the QTCP.

In a similar vein one might wonder too whether for the present spin-1 
$J_{1}$--$J_{1}'$--$J_{2}$ model there might exist a narrow strip of some 
intermediate phase, which could perhaps also act to reconcile our results 
with standard Ginzburg-Landau theory.  Again, such a possibility cannot be 
ruled out with complete certainty by any numerical calculation such as ours.  
However, we have shown that our own extrapolation schemes are sufficiently robust 
and show sufficient internal consistency to rule out any but a very narrow strip 
of an intermediate phase for $0 < J_{1}'/J_{1} \lesssim 0.66$.  We estimate that 
the width of such a strip cannot exceed by more than a factor of three or so 
that shown in fig.~\ref{spin1_phase} from the data used in the present extrapolation.  
However, we note that in the limiting case of the spin-1 1D chain (where 
$J_{1}' \rightarrow 0$, $J_{2} \rightarrow 0$) the actual GS is the Haldane gapped 
state~\cite{Ha:1983}.  Presumably this state should persist for small enough 
perturbations corresponding to small nonzero values of $J_{1}'$ and $J_{2}$.  The 
only other numerical study of the spin-1 $J_{1}$--$J_{1}'$--$J_{2}$ model of 
which we are aware~\cite{Mo:2006,Mo:2006_b} focused particular attention on 
this regime, and did indeed observe the continuation of the Haldane phase 
in a narrow strip in this regime.  Our own results are not inconsistent with 
these observations, but our interest here lies more in the case of stronger 
interchain couplings where $J_{1}'/J_{1}$ and $J_{2}/J_{1}$ are not confined 
to be small.  However, we note that other CCM calculations aimed specifically 
at this regime do, indeed, detect the Haldane gap.  Thus, Zinke 
{\it et al.}~\cite{Zi:2008} investigate the magnetic LRO of weakly coupled 
(quasi-1D) Heisenberg antiferromagnetic chains for both the spin-1/2 and spin-1 
cases, using the CCM to calculate the staggered magnetisation and its dependence 
on the interchain coupling strength ($J_{\perp}$).  They find that for the $s=1/2$ 
case an infinitesimally small $J_{\perp}$ suffices to stabilize magnetic LRO, 
whereas for the $s=1$ case a nonzero (albeit small) $J_{\perp}$ is needed to 
establish LRO, in agreement with the results from other methods.

Finally, in reaching our conclusions we have relied on two of the unique
strengths of the CCM, namely its ability to deal with highly frustrated systems
as easily as unfrustrated ones, and its use from the outset of infinite lattices.  
There is no doubt that the many-body system studied here is highly non-trivial, 
and one cannot perhaps expect any single analysis or method 
to solve it completely.  However, we present for the first time new 
and interesting results using a method for which much previous work has shown 
its ability to describe quantum phase transitions reliably.

\end{document}